\documentclass{svproc}
\usepackage{url,graphicx}

\begin{document}
\mainmatter              
\title{Bosonic drops with two- and three-body interactions close to the unitary limit}
\titlerunning{Bosonic drops}  
%
\author{A. Kievsky\inst{1} \and A. Polls\inst{2} \and
B. Juli\'a-D\'\i az\inst{2} \and N. Timofeyuk\inst{3} \and M. Gattobigio\inst{4}}
\authorrunning{A. Kievsky et al.} 
\institute{Istituto Nazionale di Fisica Nucleare, Largo Pontecorvo 3, 56100 Pisa, Italy,
\and
Departament de F\'\i sica Qu\`antica i Astrof\'\i sica, Facultat de F\'\i sica,
Universitat de Barcelona, E-08028 Barcelona, Spain,
\and
Department of Physics, University of Surrey, Guildford, Surrey GU2 7XH, United Kingdom
\and
Universit\'e C\^{o}te d'Azur, CNRS, Institut de Physique de Nice,
1361 route des Lucioles, 06560 Valbonne, France}

\maketitle              

\begin{abstract}
When the binding energy of a two-body system
goes to zero the two-body system shows a continuous scaling invariance governed
by the large value of the scattering length. In the case of three identical bosons, the
three-body system in the same limit shows the Efimov effect and the scale invariance is
broken to a discrete scale invariance. As the number of bosons increases
correlations appear between the binding energy of the few- and many-body systems. We discuss 
some of them as the relation between the saturation properties of the infinite system and the 
low-energy properties of the few-boson system.

\end{abstract}
\section{Introduction}
The ground state properties of $^4$He and $^3$He droplets with $N$ atoms
have been studied in a series of papers~\cite{kalos,usmani,pandha1,pandha2}. 
The energy per particle, $E_N/N$ can be described, as $N\rightarrow\infty$, 
using a liquid-drop formula in terms of $x=N^{-1/3}$
\begin{equation}
E_N/N=E_v + E_s x + E_c x^2
\label{eq:ldf}
\end{equation}
where $E_v$, $E_s$ and $E_c$, are the volume, surface and curvature terms respectively.
Results for the infinite liquid can be obtained from calculations at fixed
values of $N$. Since the value at saturation can be obtained independently, these
studies probe the validity of the extrapolation formulas used to predict the properties 
of the infinite system typically computed in droplets having a few hundred atoms. 

More recently helium drops have been studied
using modern helium-helium interactions~\cite{rafa,boronat}. In Ref.~\cite{lewerenz} a
diffusion Monte Carlo (DMC) method has been used to study clusters up to 10 atoms
interacting through the Tang, Toennies, and Yiu (TTY) potential~\cite{TTY}. 
Helium trimers and tetramers have been studied around the unitary limit varying the
potential strength~\cite{greene,barletta2001,hiyama1,hiyama2}. It has been shown that
with a very small reduction of the strength (about $3$\%) the binding
energy of the helium dimer disappears. In fact the helium dimer is very close to the
unitary limit having a two-body binding energy of about 1.3 mK and a large two-body
scattering length of about $189\;$a$_0$ ($a_0$ is the Bohr radius).

We can observe two, very different, descriptions of light helium clusters. On one hand, 
several models of the helium-helium interaction are available. 
On the other hand, the large value of the helium-helium scattering length locates the
small clusters of helium close to the unitary limit in which universal behavior
can be observed. Accordingly, the particular form of the potential is not important, 
many properties are determined from a few parameters as the two-body scattering length $a$ and 
the trimer ground state energy $E_3^0$ (or the first excited state $E_3^1$).
Specific (soft) potential models can be constructed in order to reproduce those
data and used to calculate binding energies of droplets and the
saturation properties of the infinite system~\cite{arturo}. In this way, 
a direct link between the low energy scale (or long-range correlations)
and the high energy energy scale (or short-range correlations) can be established. 
\section{Helium dimer and trimer with soft potential models}
In the following we study the ground state energy of the $N=2,3$ boson systems using 
a soft gaussian potential constructed  to reproduce the low-energy behavior of the
system. We define the two-body interaction as
\begin{equation}
 V(r_{ij})= V_0 e^{-r_{ij}^2/d_0^2}
\end{equation}
with the two gaussian parameters, $V_0$ and $d_0$, determined from the dimer
energy, $E_2$, and the two-body scattering length $a$. Realistic helium-helium
potentials can be used to calculate $E_2$ and $a$, subsequently used to fix $V_0$ and $d_0$.
In this way, the gaussian interaction results in a low-energy representation of the original
potential. Using the LM2M2 interaction~\cite{lm2m2}, widely used in the description of 
helium clusters,
as the reference interaction, the values $V_0=-1.2343566\,$K
and $d_0=10.0\,a_0$ can be used.
To study correlations between observables we can start analyzing the
Efimov radial law
  \begin{eqnarray}
    \label{eq:energyzrA}
      E_3^n/(\hbar^2/m a^2) = \tan^2\xi \\
      \kappa_*a = {e}^{(n-n^*)\pi/s_0} 
      \frac{{e}^{-\Delta(\xi)/2s_0}}{\cos\xi}\,,
    \label{eq:energyzrB}
  \end{eqnarray}
that gives, in the zero-range limit, the three-boson spectrum $E_3^n$ in terms of
the universal function $\Delta(\xi)$ and the three-body parameter $\kappa_*$,
defined by the energy at the unitary limit of the reference level $n^*$,
$E_3^{n^*}=\hbar^2\kappa_*^2/m$. Eq.(4) indicates that the
product $\kappa_*a$ is a function of the angle $\xi$. 
Assuming that for real systems the 
product $\kappa_*a$ is still a function of $\xi$ we can propose:
\begin{equation}
 \kappa_*a=[\kappa_*a]_G
\end{equation}
where $[\kappa_*a]_G$ is the value of the product calculated with the gaussian
potential at the angle $\xi$. To verify this hypothesis we consider the
ground state binding energies of the dimer $E_2=1.303\;$mK and trimer $E_3=126.4\;$mK as given by
the LM2M2 potential defining the angle $\xi$ as $E_3/E_2=\tan^2\xi=97.0$.
Modifying the strength of the gaussian potential to $V_0=-1.24294\,$K
and calculating the dimer and trimer energies, the same angle is obtained.
Moreover, a gaussian potential has the property that its three-body parameter
verifies $\kappa_*=0.488/d_0$~\cite{kievsky2015,raquel}. The two-body scattering length
using the modified strength is $a=170.50\;$a$_0$.
Accordingly, we can estimate the three-body parameter $\kappa_*$ of the
LM2M2 interaction, knowing that the scattering length is 189.41$\;$a$_0$, as
\begin{equation}
 [\kappa_*]_{LM2M2}=\frac{170.5}{189.41}0.488/d_0 \,\, .
\end{equation}
The obtained value is $\kappa_*=0.044\;$a$_0^{-1}$
in complete agreement with the LM2M2 value given in the literature. We have shown that
the three-body parameter can be determined by three quantities, the dimer and trimer energies
and the two-body scattering length.

\section{Saturation properties of the N-boson system}
In the following we analyze correlations between the saturation properties of the
infinite system and the low-energy behavior of the few-boson systems. To this end
we use as the reference interaction the Aziz HFDHE2 potential used in
Ref.~\cite{pandha1} to compute binding energies of helium droplets. A low energy
representation of the HFDHE2 potential is obtained by defining the parameters
of the gaussian potential $V_0=-1.208018\,$K and $d_0=10.0485\,a_0$,
giving a trimer ground state binding energy of $E_3^0=139.8\,$mK.
This value is substantially greater than the value obtained using
the HFDHE2 potential: $E_3^0=117.3\,$mK. It is well known that to tune the trimer
binding energy to the expected value
a slightly repulsive three-body force has to be introduced.
As proposed in Refs.~\cite{kievsky2011,gatto2011,gatto2012,timofeyuk1,timofeyuk2} 
we define the following three-body force
\begin{equation}
 W(\rho_{ijk})= W_0 e^{-2\rho_{ijk}^2/\rho_0^2}\,\, ,
\end{equation}
where $\rho_{ijk}$ is the hyperradius of particles $i,j,k$ defined as
$\rho^2_{ijk}=(2/3)(r^2_{ij}+r^2_{jk}+r^2_{ki})$. For selected values of
the range $\rho_0$, the strength $W_0$ is fixed to reproduce
the HFDHE2 trimer ground state binding energy $E^0_3$. The binding energy of the droplets
$E_N$ can be computed using this soft gaussian potential (SGP)
and can be studied as a function of the range $\rho_0$. In Fig.1 the binding
energies of helium drops up to $N=10$ are shown as a function of $\rho_0$
and compared to the HFDHE2 values from Ref.\cite{pandha1} using the Green Function
Monte Carlo (GFMC) method.
Though a small dependence on $\rho_0$ can be seen, an overall good description
is obtained. 

\begin{figure}[h]
\includegraphics[scale=0.35,angle=-90]{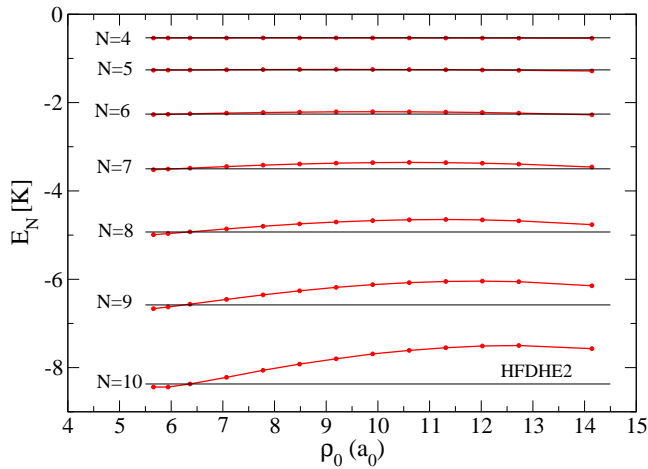}
\caption{Binding energy per particle up to $N=10$ atoms.
The results of the SGP for different values of the three-body
force range $\rho_0$ are shown as red circles and compared to
the HFDHE2 results.}
\label{fig1}
\end{figure}

The $\rho_0$ dependence is analyzed in Fig.2 in the case of the tetramer binding
energy. It can be seen that there is a value of $\rho_0$, around
8.5$\;$a$_0$, that gives the best description of this quantity. The next step is to
compute the droplets binding energies up to $N\approx 100$ and extract the saturation
energy from Eq.(1). This is shown in Fig.3 where the results
for different values of $\rho_0$ form
the dark band. The results using the optimum value of $\rho_0=8.5\;$a$_0$
are shown as (blue) points. They follow, with acceptable accuracy, the
GFMC results using the HFDHE2 potential shown as the (red) solid line.
Using the optimum value of $\rho_0$  it is possible to determine 
$E_v$, $E_s$ and $E_c$ defined
in Eq.~(\ref{eq:ldf}).
From the results of the SGP in the range $20\le N \le 112$ 
the following values are obtained (in K)
\begin{equation}
E_N/N= 6.98 - 18.6\, x + 10.3\, x^2 \,\,\, .
\label{eq:ldf1}
\end{equation}
They can be compared to the values (in K) obtained 
with the GFMC method $E_v=7.02$, $E_s=-18.8$ and $E_c=11.2$
using the HFDHE2 interaction.

\begin{figure}[h]
\includegraphics[scale=0.35,angle=-90]{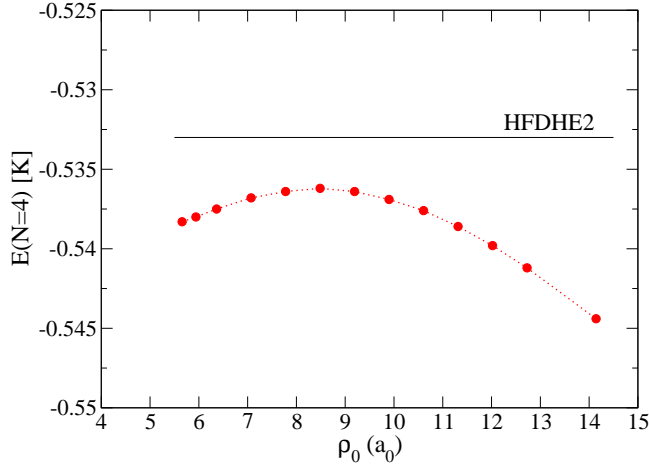}
\caption{The tetramer binding energy as a function of the three-body
force range $\rho_0$. The value of
the HFDHE2 potential is shown.}
\label{fig2}
\end{figure}

We conclude that after tuning the range of the three-body force
to reproduce as better as possible the tetramer binding energy, the
soft gaussian potential, consisting of a two- and a three-body term,
with the four parameters determined by the dimer, trimer and tetramer
binding energies and the two-body scattering length is able to
estimate with good accuracy the energy per particle of the
infinite system.
\begin{figure}[h]
\includegraphics[scale=0.38,angle=-90]{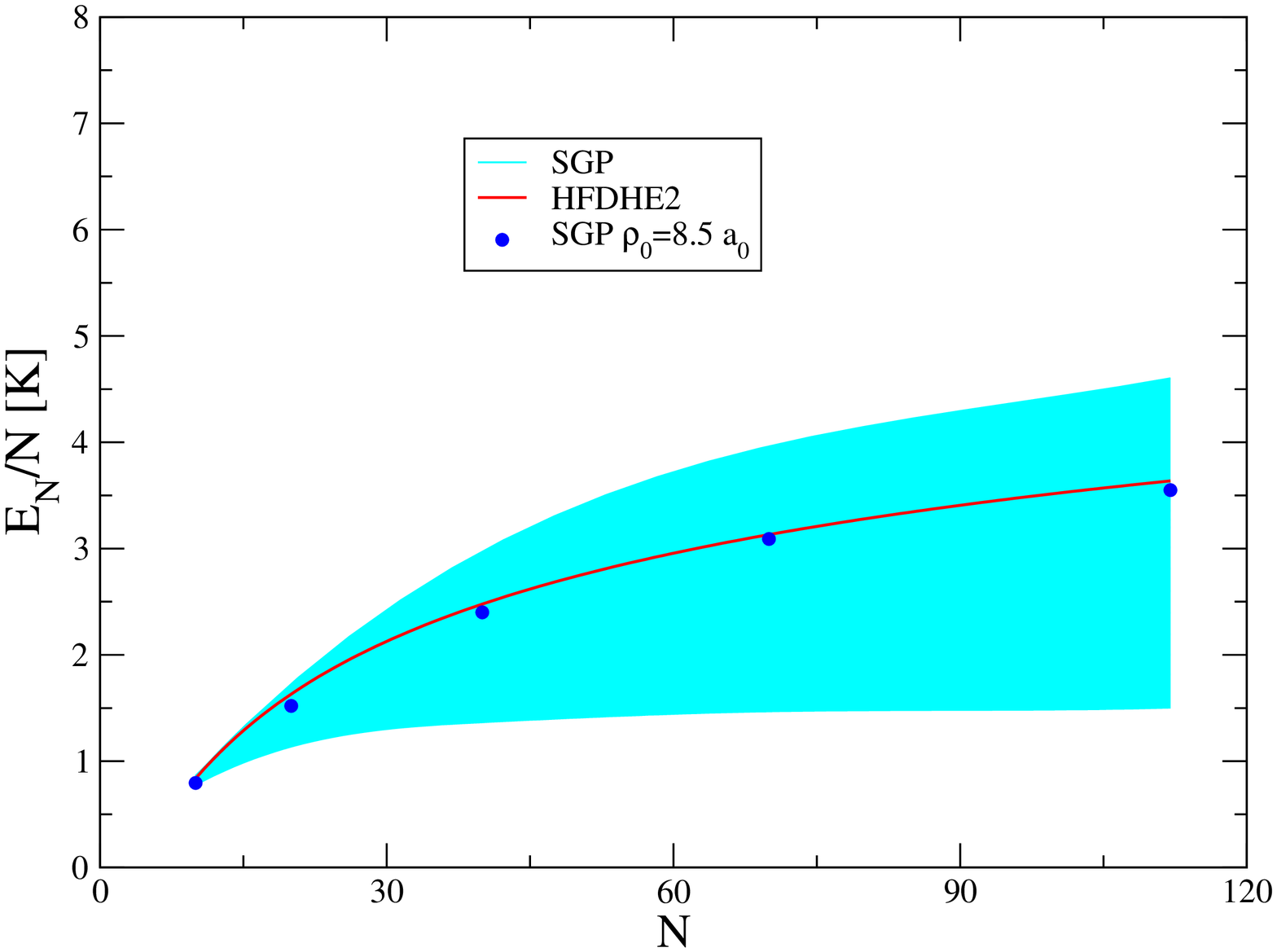}
\caption{The binding energy of the droplets as a function of the
number of particles $N$}
\label{fig3}
\end{figure}
\section{Conclusions}

In the present work we have analyzed correlations between different observables
imposed by the proximity of the system to the unitary limit.
Due to the large value of the two-body scattering length, helium drops
are well suited to study these phenomena. 
Correlations of this type can also be studied in nuclear systems,
since the $n-n$ and $n-p$ scattering lengths are large~\cite{kievsky2017,kievsky2018}.
Here we have shown results for helium drops using a gaussian soft interaction 
to determine the three-body parameter $\kappa_*$. Noticeably,
the result was in extremely good agreement with the values given in
the literature calculated directly using the LM2M2 potential.
Secondly, using the HFDHE2 as the reference potential, we have calculated
binding energies for helium drops up to $N=112$ and, using a liquid-drop
formula, we have extracted the saturation energy. We have observed that
using the optimum value for the range of the three-body interaction
a good estimate of the experimental saturation energy is
obtained. In this way we have clarified the existing correlations between
different observables imposed
by the unitary limit in many-body systems close to the unitary limit.

%
%

\end{document}